# Calendar Time Local Earthquake Forecasts from Earthquake Nowcasts: A Do-It-Yourself (DIY) Ensemble Method

John B Rundle[1,2,3], Ian Baughman[1], Andrea Donnellan[4,3], Lisa Grant Ludwig[5], Geoffrey Fox[6], Kazuyoshi Nanjo[7]

[1] University of California, Davis, CA
[2] Santa Fe Institute, Santa Fe, NM
[3] Jet Propulsion Laboratory, Pasadena, CA
[4] Purdue University, West Lafayette, IN
[5] University of California, Irvine, CA
[6] University of Virginia, Charlottesville, VA
[7]Shizuoka University, Japan

## Abstract

This paper presents a new technical method for computing calendar time forecasts in a local area for large earthquakes of a target magnitude $M_T$ using a count small earthquakes $M_S < M_T$ in the area, together with the Gutenberg-Richter (GR) magnitude-frequency relation. The GR relation states that for every large target earthquake of magnitude greater than $M_T$ , there are on average $N_{GR}$ small earthquakes of magnitude $M_T >$M $\geq M_S$. The only assumption is that the GR statistics of the local area are the same as in the larger surrounding regions. This assumption is used to construct an ensemble of earthquakes in larger surrounding regions to be used in computing the forecast. The method has significant skill, as defined by the Receiver Operating Characteristic (ROC) test, which improves as time since the last major earthquake increases. The probability is conditioned on the number of small earthquakes $n(t)$ that have occurred since the last large earthquake. There is no need to assume a probability model, as the probability is instead computed directly as the Positive Predictive Value (PPV) associated with the ROC curve. The method is validated by comparison to the UCERF3 forecasts for the UCERF3-defined geographic boxes centered on Los Angeles and San Francisco. The method is then applied to a 125-KM radius circular area around Los Angeles, California, following the January 17, 1994 magnitude $M$6.7 Northridge earthquake, and short term forecasts (1 year and 5 year ) are computed.

## Key Points

- Earthquake nowcasting uses counts of small earthquakes to determine the current real time probability of a large earthquake
- The technique can be used as a basis for earthquake forecasting in calendar time
- A major advantage is that the forecast probability curve is determined directly from the data itself, rather than being assumed

## Plain Language Summary

Earthquake nowcasting is a recently developed method that allows the current progress of a seismically active region through its cycle of large damaging earthquakes to be determined. Forecasting and predicting the location and time of major earthquakes are long-sought goals. Unlike weather forecasting, the data needed for detailed and precise earthquake forecasting will always be incomplete. Among data that can be observed are earthquake catalogs, which include the time, location, and magnitude of the events. In past works, we have shown that the current state of the local regions can be observed by counting the number of small earthquakes since the last large earthquake in a region. We have called this method earthquake nowcasting. In a very recent paper, we developed a new technique for extending earthquake nowcasts to earthquake forecasts in the "natural time" domain, where by natural time we mean counts of successive small earthquakes. In the present paper, we extend the method to the calendar time domain, thus answering the question, " what is the probability of a large earthquake of magnitude X in a small region during the next Y months?" We then illustrate these procedures by application to the region around Los Angeles, CA following the January 17, 1994 magnitude M6.7 Northridge earthquake. Major advantages of the proposed method are its basic simplicity, and the fact that there are no unknown parameters that must be assumed or set by arbitrary means.

## Introduction

Forecasting and predicting the location and time of major earthquakes are long-sought goals. Unlike weather forecasting, the data needed for detailed and precise earthquake forecasting will always be incomplete. A relatively recent description of the state of earthquake forecasting is described in, for example, Jordan et al. (2011) and Rundle et al. (2021a).

Here we describe a new technical method based only on the statistical analysis of earthquake catalogs, with no need for fixing free parameters. In this approach, a statistical ensemble method is defined that assumes only that the ensemble of systems all obey the same Gutenberg-Richter statistics, and are thus scale-invariant. A necessary assumption in the model is that the small earthquakes used are constrained by the completeness and quality of the earthquake catalog. A bad catalog would produce a bad forecast.

The basic idea is to focus on a small seismically active local region, such as the circular region surrounding Los Angeles, CA in Figure 1 (left). We then construct an ensemble of rectangular regions encompassing the circle as shown. The large event to be forecast has a magnitude $M \geq M_T$, where $M_T$ is a target magnitude. For purposes of this paper, we assign $M_T = 6.0$. In the catalog, there are many smaller magnitude events $M_S \leq M < M_T$ where $M_S$ can be considered the catalog completeness level. Again, for the purposes of this paper, we assign $M_S = 3.5$, and the radius of the circle we use equals 125 KM.

As shown in Table 1, the circular region has many fewer events than the number in the expanding series of rectangular ensembles. In the ensemble method we introduce, the only assumption we make is that the magnitude-frequency statistics of the events in the ensembles are the same as the statistics in the circle. More specifically, we can use ratio of the total number of events in an ensemble, and the total number of event in the circle, to time-scale calendar times of the events in the ensemble member. Or we can time-scale the forecast time interval that has been selected for the circular region. This process is discussed below.

Each ensemble consists of a number of cycles activity of small events $M<M_T$ between large earthquakes $M \geq M_T$. This is again compiled in Table 1 for several ensemble members. Figure 1 illustrates these cycles as time series at right. Note that in Figure 1, the time series amplitudes are constructed using the Poisson Cumulative Distribution Function for the ensemble member, as will be discussed below. After scaling the calendar forecast times in the ensemble members, we then consider each cycle of activity in a given ensemble. Collections of time-scaled cycles of activity that have fewer small events than the number in the circle are considered to represent a set of might-have-been past histories for activity in the circle. Collections of time-scaled cycles of activity that have more small events than the number in the circle are considered to represent a set of possible future activity for the circle.

With this assumption of scaled similarity, we then have a catalog of statistically possible futures for the circle. By taking ensemble averages over these possible futures, forecast probabilities can be calculated. It is for this reason that we do not need to assume any kind of statistical model with parameters that must be fixed by training on a separate data set. There is then no need for a training data set distinct from a test data set, or for out of sample testing. Validation can be carried out by comparison to other a standard forecasts such as the UCERF2 and UCERF3 30-year forecasts for California (Field et al., 2009, 2015) as we show below.

Previous papers have developed several techniques for earthquake nowcasting, the precursor to the current forecasting, which is based on using counts of small earthquakes to track the progression of the fault system through its cycle of large earthquake activity. The term nowcasting is used in the same sense as for weather and economic nowcasting (e.g., Rundle et al., 2016; Rundle et al., 2021a), the determination of the current state of a system in the recent past, current time, and the near future. Methods to produce earthquake nowcasts have been the subject of previous papers, a selection of which are: (Rundle et al., 2016; 2018; 2019a; 2019b; 2020; 2021a,b; 2022a,b; 2024; Pasari and Mehta, 2018; Pasari, 2019; Pasari, 2020; Pasari and Sharma, 2020; Chouliaras, 2009; Chouliaras et al., 2023; Perez-Oregon, 2020).

## Receiver Operating Characteristic Analysis

The most recent paper (Rundle et al., 2026) develops a technique to compute the probability of a future large earthquake directly from the catalog data, by the use of the Receiver Operating Characteristic (ROC). The ROC method, developed in the 1940's by the British with the advent of radar (Junge and Dettori, 2018), considers all possible cases of a signal followed by an event:

- True positive (TP), where a signal is observed, followed by an event
- True negative (TN) where no signal is observed and no event follows
- False positive (FP), where a signal is observed and no event follows
- False negative (FN), where no signal is observed but an event follows

These are the basic quantities that we use in building the forecast.

The nowcasting method is applied to a small circle of a chosen radius surrounding a point of interest, typically a city. Computing the probability then involves collecting statistical data using an expanding series of larger rectangular regions surrounding the circle. These regions then comprise the ensemble set for the method. More specifically, the analysis involves computation of the ROC curve, a plot of the True Positive Rate (TPR) vs. the False Positive Rate (FPR), where these are defined as e.g., (Mandrekar, 2010; Powers, 2011):

$$TPR = TP/(TP + FN) \quad (1)$$

$$FPR = FP/(FP + TN)$$

Once the ROC curve is computed for a member of the ensemble, the Positive Predictive Value (PPV) is then computed:

$$PPV = TP/(TP + FP) \quad (2)$$

The basic assumption in this method is that the catalog statistics of the surrounding region, essentially the Gutenberg-Richter *b*-value, are the same as the statistics of the circular region at a 95% confidence level. Or more specifically, we make what amounts to an ergodic assumption, that time averages can be replaced by space (or ensemble) averages.

This process produces a family of ROC and PPV curves, one corresponding to each member of the ensemble. Once we are in possession of these curves, we compute the mean curve and its standard deviation. Note that the area under the ROC curve (AUROC, or "skill") can be interpreted as the ability of a model to discriminate between, or to correctly classify, different categories of events. In our case, an event is the occurrence of a major earthquake vs. non-occurrence (Mandrekar, 2010; Powers, 2011).

## Ensemble Construction

In Rundle et al. (2025), we used a fixed number of hypothesized future small earthquakes for our forecast interval. Counts of small earthquakes are an example of "natural time", the time scale that is relevant to the physics of the system. But these small earthquakes occur with unknown future frequency and timing. So a fixed future count is not as informative as it could be. Since human beings operate on calendar time, forecasts are most useful if a fixed future time interval is used.

This problem motivates the use of the ensemble approach. To implement this approach, we first define the minimum rectangular ensemble surround the circle of interest. We characterize this size as a rectangle with half width of size $D_{M.}$ The criterion we use for the minimum rectangle size is to require a minimum of 20 large earthquakes of a target magnitude $M_T$. "Small earthquakes" are defined as events having magnitude $M_S < M_T$.

For each member of the ensemble, we classify "cycles of activity" between large events whose beginnings and endings are bounded by earthquakes of the target magnitude $M_T$. So an ensemble member with $N_{ENS}$ large earthquakes will have $N_{ENS}$ -1 cycles by this definition. Cycle *j* within ensemble member *i* ($C_{i,j}$) will have $n_{i,j}$ small earthquakes by definition.

Additional members of the expanding ensemble are then considered as those with an increasing half width of value $\Delta D$. For the example we show below, we require a minimum of 20 earthquake cycles for statistical validity, so we find that $D_M = 3.6^o$. We then define an additional 29 members of the ensemble as those with $D = 3.7^o, 3.8^o, 3.9^o$, etc. thus $\Delta D = 0.1^o$, for a total number of ensemble members = 30. See Table 1.

A critically important point is to note that these increasingly large rectangular regions contain many more small earthquakes over the same time period than the small circular region in which we wish to make the forecast. So in terms of "calendar time", "natural time" will pass increasingly more quickly in the successively larger ensemble members than it does in the small circular region.

Or to state the idea in another way, the rate of occurrence of events in a larger region $R_L$ is higher than the rate in the circle $R_C$, $R_L > R_C$, so the natural time "clock" is running faster in the larger region than in the circle (Table 1). As a result we need to apply a correction for this effect.

In Rundle et al. (2026), we used a single optimized large region surrounding the circle. We then applied a correction to the calendar time rate in the large region to match the rate in the circle. In the present application in which we use an ensemble of large regions, we are using a future forecast time interval $T_F$ rather than a forecast number. So rather than scale the regions in time, we choose to scale the forecast time to a value appropriate for each large region in the ensemble:

$$T_{F,L} = (R_C / R_L) T_{F,C} \tag{3}$$

Here $T_{F,C}$ is the chosen forecast time of interest in the circle, and $T_{F,L}$ is the corresponding scaled forecast time in the region. Clearly, $T_{F,C} > T_{F,L}$ since $R_L > R_C$. As a result, each region in the ensemble will have a different scaled forecast time, the scaled forecast time decreasing as the linear dimension of the region increases.

## Building the ROC Curve

To calculate the forecast probability for a fixed future time interval $T_{F,C}$ in the circle of interest, we build the ROC curve by considering a sequence of small earthquakes in the circle following the last large earthquake. These small earthquakes in the circle catalog occur at a sequence of times that we denote by $t_{n_C}$, where $n_C = 1,...,N_C$, which are known from the circle catalog. Similarly, in cycle $j$ of ensemble $i$, denoted by $C_{i,j}$, there is some distinct number of small earthquakes $N_{i,j}$.

In a given cycle, there may be almost no small earthquakes where mainshocks cluster strongly in calendar time. In other cycles, there may be many small earthquakes where there is a comparatively large separation between mainshocks in calendar time.

Now consider a cycle $C_{i,j}$ having a total number $N_{i,j}$ of small earthquakes. As in Rundle et al. (2026), we define a nowcast value $\Phi$ by:

$$\Phi(n) = 1 - e^{-(n/N_{scale}))} \tag{4}$$

where $N_{scale}$ is computed from the Gutenberg-Richter relation (e.g., Boore, 1989):

$$N_{scale} = e^{b(M_T - M_S)} \tag{5}$$

Here $b$ is the Gutenberg-Richter $b$-value for the ensemble member. Equations (4)-(5) are equivalent to the Earthquake Potential State (EPS) as defined in (Rundle et al., (2016; 2018;

2019a; 2019b; 2020; 2021a,b; 2022; 2024). In addition, $n$ in equation (4) represents all events contained in an ensemble member.

These equations arise from the idea that the interval statistics of large mainshock earthquakes, excluding aftershocks are generally Poisson distributed (e.g., Gardner and Knopoff, 1974), as shown in Figure 1. Applying equations (4) and (5), we show a typical ensemble of nowcast timeseries in Figure 2 (right) for a region surrounding Los Angeles, CA, which plots nowcast values as a function of time.

We first build the ROC using all cycles in the ensemble, then apply to the circle. Using (4), all times in the ensemble of cycles are assigned nowcast values $\Phi([n_{i,j}])$ where $[n_{i,j}]$ is the set of sequence indices of the small earthquakes in the $j^{th}$ cycle of the $i^{th}$ ensemble member. As an explicit example, if there are 5 small earthquakes in cycle 3 of ensemble member 2, $[n_{2,3}] \in [1,2,3,4,5]$. Then $\Phi([n_{2,3}])$ would be a set of 5 values computed by applying equations (4) and (5) to the set $[n_{2,3}]$, yielding the set $[\Phi_1, \Phi_2, \Phi_3, \Phi_4, \Phi_5]$. Note that each of the events in $C_{i,j}$ has a time value as well as a sequence number, $t \in [t_1, t_2, t_3, t_4, t_5]$.

So the critical point is that each small event in the ensemble can be characterized by a nowcast value, a sequence index, and a time value. To build the ROC diagram, we adopt a set of threshold values, as has been explained in, e.g, Rundle et al. (2021a,b; 2022a,b; 2024) . For each ensemble member, we classify all points on the nowcast timeseries.

We start by selecting an arbitrary threshold value $\tau$. Given a small event occurring at time $t$, we ask if that event's nowcast value is above or below the threshold. We also ask if the next target earthquake $M_T$ occurs after time $t$ but within the time interval $t + T_{F,L}$, where again, $T_{F,L}$ is the scaled forecast time for that ensemble member. Classification is then given by:

- TP, if the nowcast value is above the threshold $\tau$, and the next target earthquake occurs within $t + T_{F,L}$
- FP, if the nowcast value is above the threshold $\tau$, and no earthquake occurs within interval $t + T_{F,L}$
- FN, if the nowcast value is below the threshold $\tau$, and the next target earthquake occurs within interval $t + T_{F,L}$
- TN, if the nowcast value is below the threshold $\tau$, and no target earthquake occurs within interval $t + T_{F,L}$

It should be noted that the ROC diagram is conditioned on the current number of events in the circle. For example, if the current number of small earthquakes in the circle is, for example 100, then no cycles of length less than 100 are used in the classification. Examples of conditional ROC diagrams will be shown below.

We should also explicitly note that the various computed quantities TP, TN, FP, FN are functions only of the threshold values, $\tau$, so that we have TP($\tau$), TN($\tau$), FP($\tau$), FN($\tau$). We use the cycles in each ensemble to build an ROC curve for that ensemble. Then all cycles are plotted as in Figure 3, and the mean of those curves is found.

## Computing the PPV Probability

To compute the PPV values (probabilities) at the times of the small events following the last large event in the circle, we use as threshold values $\tau$ the nowcast values for these small events defined by (4) and (5). So if there are 100 small earthquakes within the circle since the last large earthquake, we compute 100 values for $\Phi(n)$ and use these as our threshold set $[\tau(n)] \equiv [\Phi(n)]$.

Each of these small earthquakes then has a defined nowcast value, a sequence number, and an occurrence time, all of which are associated with the event. Again, note that the TP, FP, TN, FN are functions of the threshold only, which in this special case is the event nowcast value. Since the nowcast value is also a threshold, this association allows to identify specific event sequence numbers and event times with a PPV value.

Continuing the example in the previous section, we would then have a set of defined thresholds $[\tau] = [\Phi_1, \Phi_2, \Phi_3, \Phi_4, \Phi_5]$ with associated times $t \in [t_1, t_2, t_3, t_4, t_5]$. This allows us to associate the threshold values, which the TP, FP, TN, FN depend on, with event occurrence times in the circle.

## Results

We apply our method to a circular region of radius 125 km surrounding Los Angeles, CA, using a target magnitude $M_T$ = 6 and small earthquake $M_S$ = 3.49. The most recent large earthquake in the circle was the Northridge earthquake, having magnitude $M$=6.7, occurring on January 17, 1994.

The choice of circle radius is arbitrary, but we use a rule of thumb that the radius should be a 2-3 times the linear dimension of the aftershock zone of the previous large earthquake in the circle, which in this case is the Northridge event. In addition, Chouliaras et al. (2023) have conducted a more quantitative analysis of appropriate circle radius for similar earthquake magnitudes in Greece, and also arrived at a radius of 125 km. Note that the forecast here applies only to the Northridge event.

An additional criterion for our choice of radius is the "radius of significant ground shaking" for a given magnitude, for example the distance at which one might experience Mercalli Intensity VI (PGA $\sim 0.1g$) shaking. As Manyele and Mwambela (2014), Minson et al. (2021), and many others have shown, this distance is about 100-150 km for an $M_T$ = 6 earthquake which is our minimum size for the forecast.

In Figure 3, we show a plot of 4 conditional ROC diagrams for a 1-year forecast assuming 1) no events have yet occurred in the circle; 2) 100 events have occurred in the circle; 3) 250 events have occurred in the circle; and 4) 448 events have occurred, the current number as of this writing. Notice that the area under the curve, or skill, increases later in the cycle. This is a result of more small events occurring, thus there are fewer and fewer cycles used to construct the ROC curve, and these increasingly do not include the initial clustered events.

The skill represents the ability of a forecast to discriminate among different cases: 1) future event occurring, or 2) no future event occurring. It can be seen that there is essentially no skill if the current number of circle events is small, but that the skill increases as more small events occur. Also note that the ROC diagrams in Figure 2 were constructed using a set of thresholds defined at 100 equidistant values between [0,1] ( or [0, 100%] ).

Also, for the purposes of these plots, we have used an ensemble of 30 regions, from 3.6° to 6.5° linear half-length as shown in the figures. The red curve represents the mean value of the ROC for the ensemble, and the dashed lines represent the standard deviations. The cyan curves are the ROC values for the various members of the ensemble.

Figure 4 shows PPV plots for 1 year and 5 years for events in the circle. For these plots, we set the threshold $[\tau] = [\Phi_{Circle}]$, where $\Phi_{Circle}$ represents the nowcast values computed for the specific events and times in the circle. Again, the red curve represents the mean, the dashed curves are the standard deviations, and the cyan curves are the PPV curves for the various ensemble members.

The figures at left are for a forecast time of 1 year, and the figures at the right are for a forecast time of 5 years. The figures at top use an ensemble of 30 regions, and the plots at bottom use an ensemble of 60 regions. These figures are a sensitivity analysis that shows that the mean PPC curve is relatively insensitive to the number of regions in the ensemble.

It can be seen in Figure 4 that for the shorter forecast intervals, mainshock clustering dominates, i.e., initial high PPV values immediately following the last (Northridge) mainshock. In the following years, the PPV values gradually increase as the buildup to the next large earthquake occurs. For the longer forecast times, the mainshock clustering effect is increasingly less important.

## Validating the Method

An important consideration is to validate the method by comparison to a recognized, accepted forecast or benchmark. The obvious choices are the UCERF2 time-independent forecast, and the UCERF3 time-dependent forecast (Field et al., 2009, 2014). Figure 5 shows the comparison. The UCERF2 forecast for magnitude 6.7, produced in 2007, computed a 30 year forecast of 67% for the Los Angeles box, a rectangular region that we show in the new Figure 5, and a 63% for the San Francisco Bay area for that rectangular box. In 2016, the UCERF3 forecast published the 30 year forecast of 60% for the same Los Angeles box. The UCERF2 30-yr forecast is especially relevant because we are past the midpoint of the 30-year time window. The UCERF3 time-dependent forecast is a valid comparison because it is the official USGS forecast.

The UCERF3 forecast was produced in 2014, whereas the ensemble forecast was produced now. But the numbers are similar, and within the quoted error bounds for the ensemble method. This similarity provides a useful benchmark for evaluating the quality of the ensemble forecast.

## Concluding Remarks

We have developed a new method for earthquake forecasting for arbitrary forecast times and for small geographic regions encompassing multiple earthquake faults. It is generally not as appropriate for individual faults, where quiescence may be the normal mode of behavior. The method is appropriate for use when a forecast for such small regions is desired on a frequent basis, and can be carried out rapidly and easily on a laptop computer in a few minutes of computation. For that reason, the ensemble method constitutes a reasonable addition to present forecast methods.

**Acknowledgements**. Research by JBR was supported in part by a grant from the Southern California Earthquake Center grant #SCON-00007927 to UC Davis, and by the John LaBrecque fund, a generous gift from John LaBrecque to the University of California, Davis. SCEC contribution number 15007**.** The authors would also like to acknowledge an informative conversation with Jeanne Hardebeck of the USGS.

**Open Research.** Python code that can be used to reproduce the results of this paper for the circular region can be found at the Zenodo site: https://doi.org/10.5281/zenodo.17871110. Python code that can be used to reproduce the results of this paper for the rectangular region can be found at the Zenodo site: https://doi.org/10.5281/zenodo.19699380.

**Data.** Data for this paper was downloaded from the USGS earthquake catalog for California, and are freely available there. An included method in the Python code mentioned above can be used to download these data for analysis.

## Tables

**Table 1.** Statistical data for selected ensemble members for cycles between successive magnitude $M \geq 6$ earthquakes from 1980.0-present. Min/Max Cycle Length is the minimum/maximum number of small earthquakes among the cycles for that region size. Total Small EQ entry is the total number of small earthquakes in the catalog for the circle or ensemble member since 1980 for the range $M \geq 3.5$. Least squares fit range for $b$-values are between $M = 3.5$ and $M = 6.5$.

| | | Ensemble Number | | | | | | |
|---|---|---|---|---|---|---|---|---|
| | | 1 | 5 | 10 | 15 | 20 | 25 | 30 |
| Region Size | 125 Km LA Circle | 3.6° x 3.6° Rectangle | 4.0° x 4.0° Rectangle | 4.5° x 4.5° Rectangle | 5.0° x 5.0° Rectangle | 5.5° x 5.5° Rectangle | 6.0° x 6.0° Rectangle | 6.5° x 6.5° Rectangle |
| Total Small EQ | 600 | 6286 | 6657 | 7345 | 7803 | 8026 | 8192 | 8777 |
| Number Cycles | - | 22 | 24 | 27 | 29 | 32 | 33 | 41 |
| Min Cycle Length | - | 15 | 10 | 10 | 10 | 2 | 6 | 7 |
| Max Cycle Length | - | 657 | 797 | 800 | 825 | 730 | 755 | 798 |
| b-value | 0.93 ± .02 | 0.96 ± .01 | 0.96 ± .01 | 0.97 ± .01 | 0.98 ± .01 | 0.96 ± .01 | 0.97 ± .01 | 0.94 ± .01 |

**Figure 1.** Diagram of the meaning of an ensemble of regions, as discussed in the text. The time series amplitude are created by inserting the accumulating small earthquakes in each cycle into the Poisson Cumulative Distribution Function, equations (3) and (4) in the text.

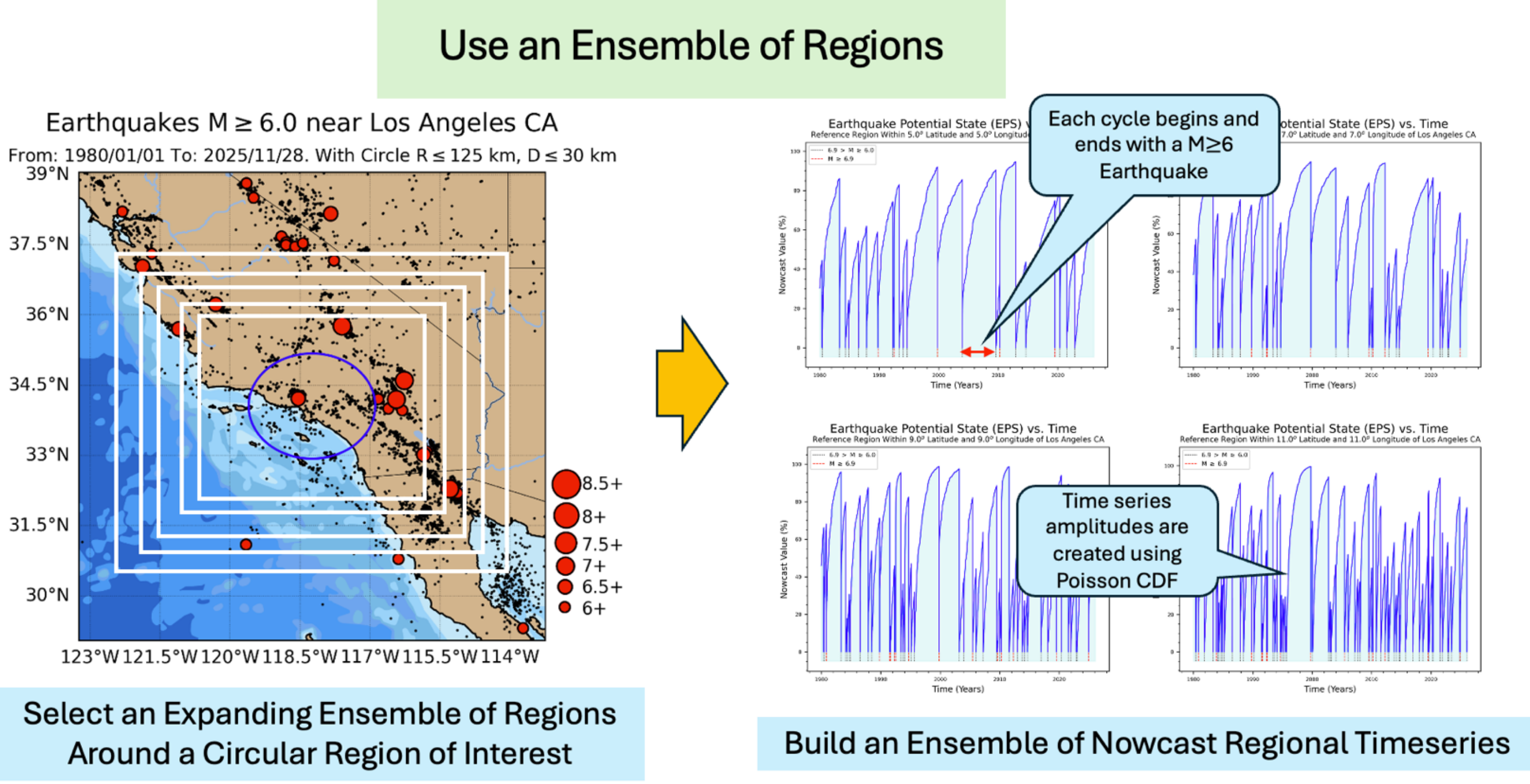

**Figure 2.** Left: Map of the greater California region, with large earthquakes shown as red circles. Blue circle is the Los Angeles area of interest. Right: Cumulative Distribution Function (red stair step curve with magenta dashed error bounds), along with the best fit Poisson CDF, as defined in equation (4).

**To Begin:**

1. Pick a region of interest: A 125 KM circle around Los Angeles
2. Plot the Histogram for the Number of Small Earthquakes $3.5 \leq M < 6.0$ Between Large $M \geq 6.0$ Earthquakes

**Cumulative Distribution Function:**

Nowcast Function for Earthquakes $M \geq 6.0$ in the Map Region

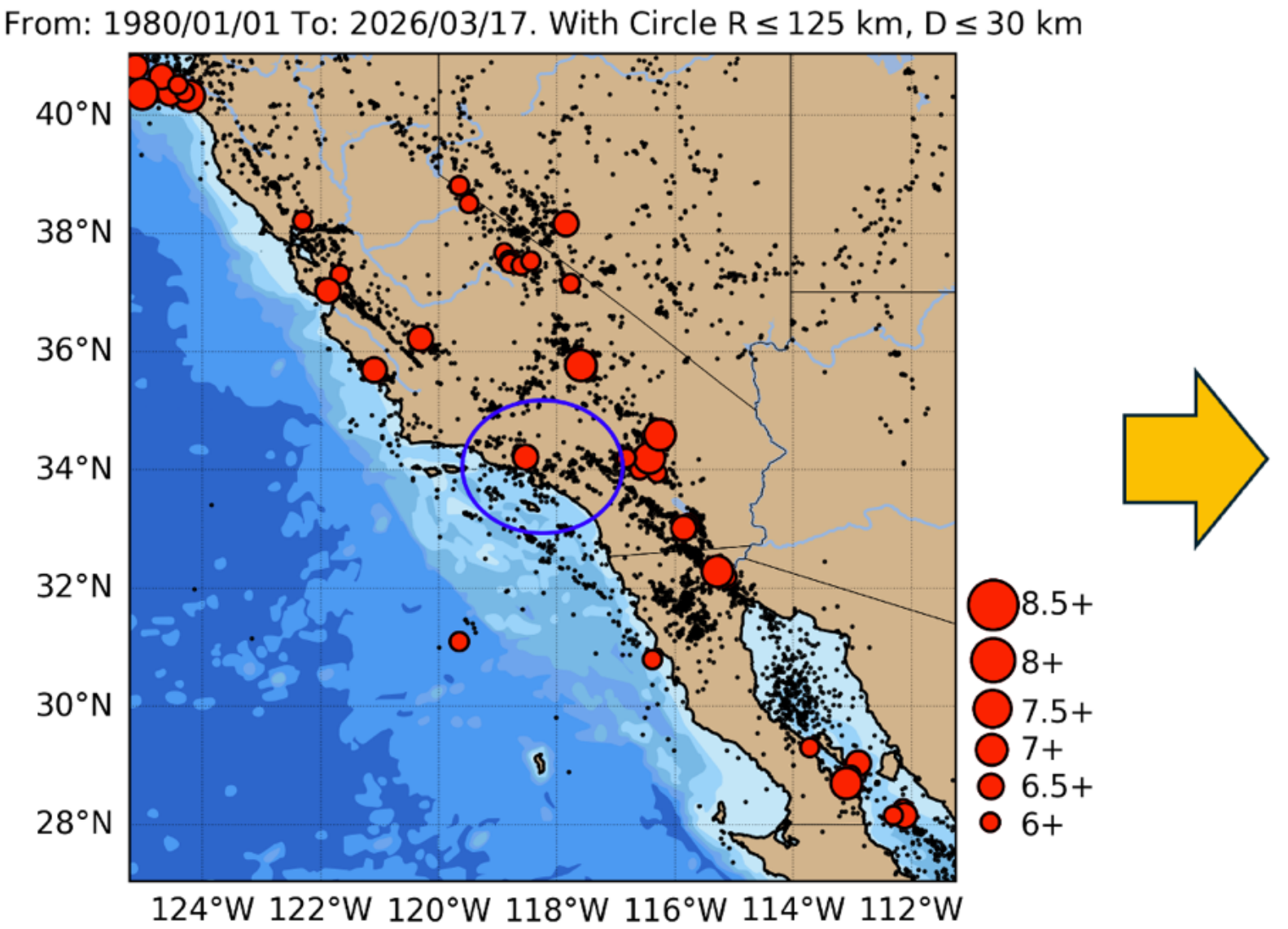

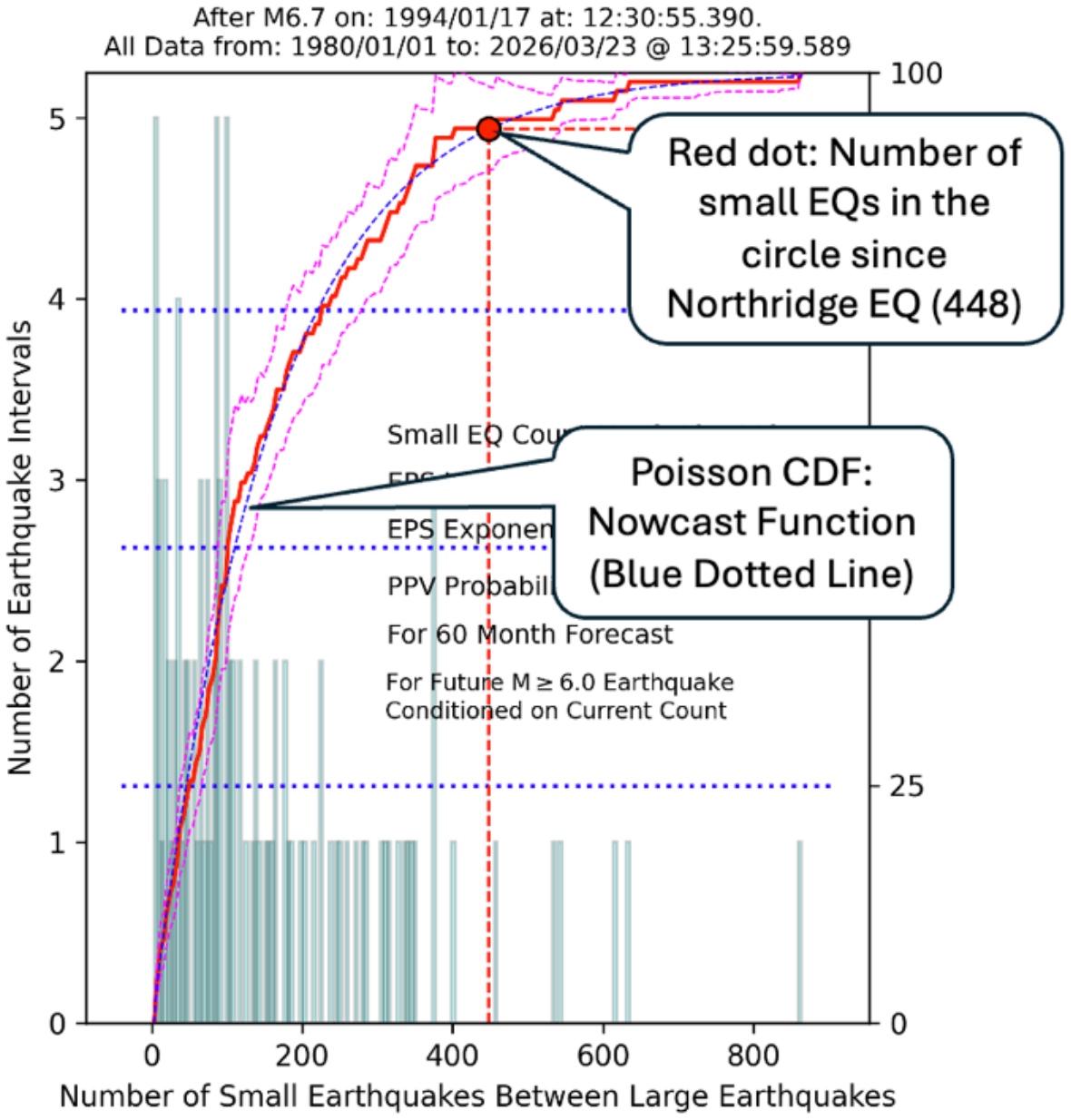

**Figure 3.** Conditional Receiver Operating Characteristic (ROC) diagrams for an ensemble of 30 regions from 3.6° to 6.5° surrounding Los Angeles, CA, at 0.1° interval half-widths, with a forecast time of $T_F$ = 5 years. The diagrams are conditional because only cycles with more events than the designated number (0, 150, 300, 448) of events are used to compute the ROC curves. Cyan curves are for the various ensemble members. Red curves are the mean values, dashed curves are the 1 standard deviation curves. Diagonal blue line is the no skill line. Skill for the curves are, respectively, 0.47, 0.78, 0.86, 0.90, showing that skill improves later in the earthquake cycle.

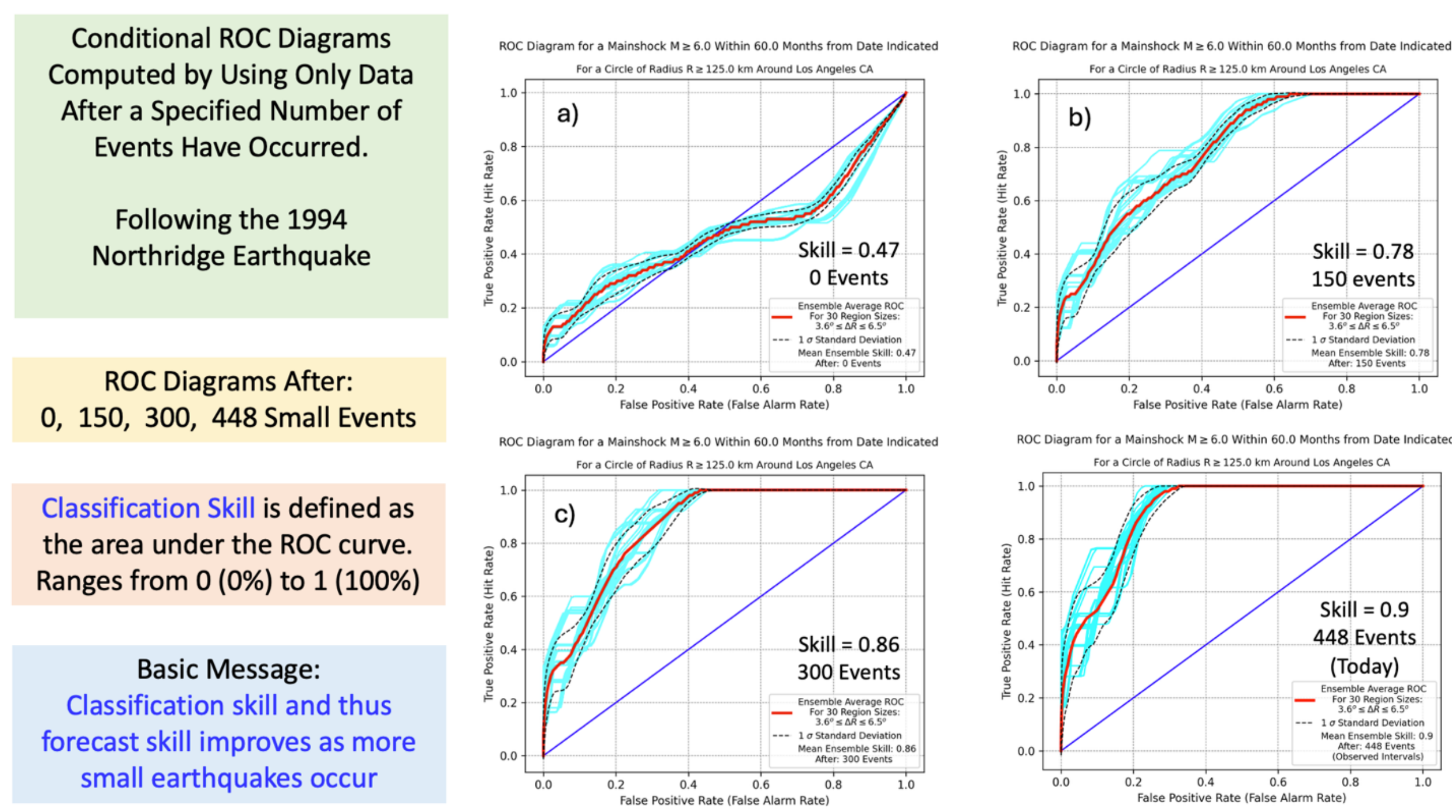

**Figure 4.** Plots of PPV, the probability of a future $M \geq 6$ earthquake, as function of time since the M6.7 Northridge, CA earthquake on 1/17/1994. a) Ensemble size = 30, Forecast time interval $T_F$ = 1 year. b) Ensemble size = 60, Forecast time interval $T_F$ = 1 year. c) Ensemble size = 30, Forecast time interval $T_F$ = 5 years. d) Ensemble size = 60, Forecast time interval $T_F$ = 5 years. Cyan-colored curves are for the various ensemble members. Red curve is the mean ensemble probability, and dashed curves are the 1-standard deviation curves.

ROC Analysis is Used to Compute Conditional Probabilities (PPV) Near Los Angeles.

Following the 1994 Northridge Earthquake

Probability (PPV) Diagrams for:

12 Month Forecasts in 2 Ensemble Regions (left)

60 Month Forecasts in the Same 2 Ensemble Regions (right)

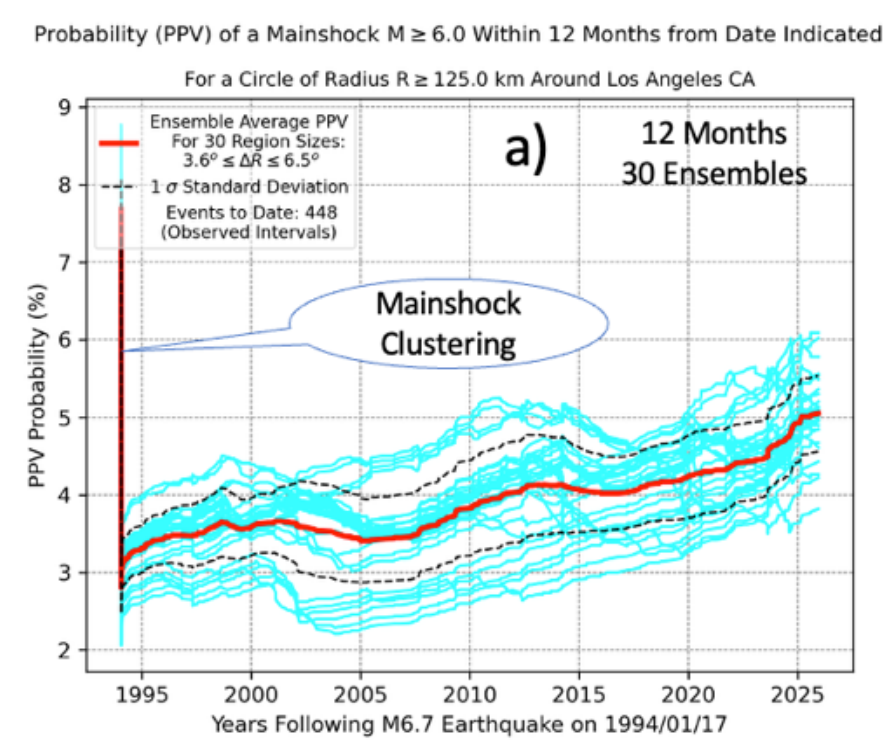

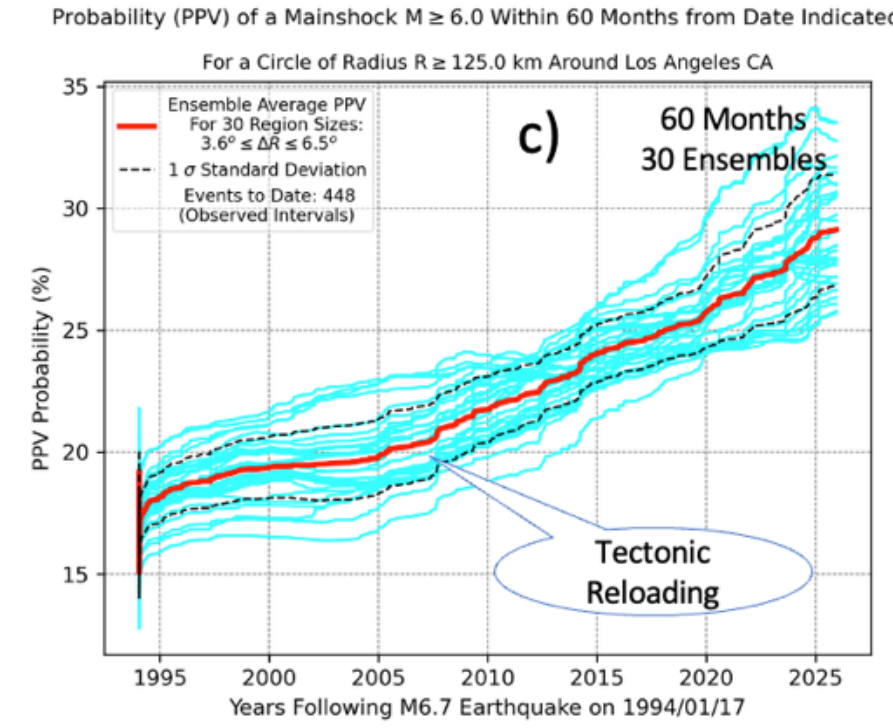

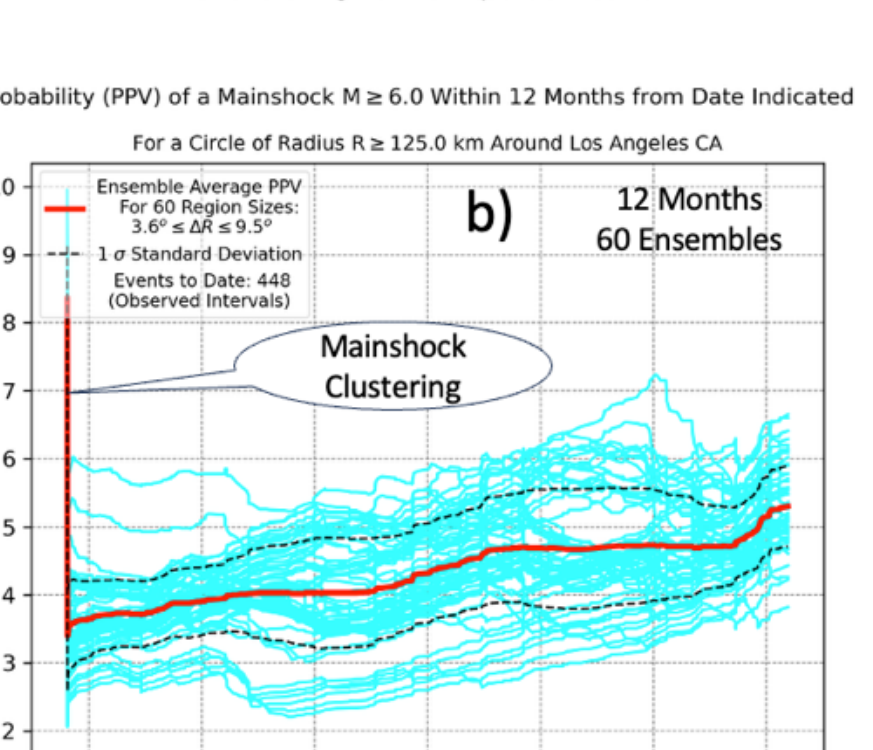

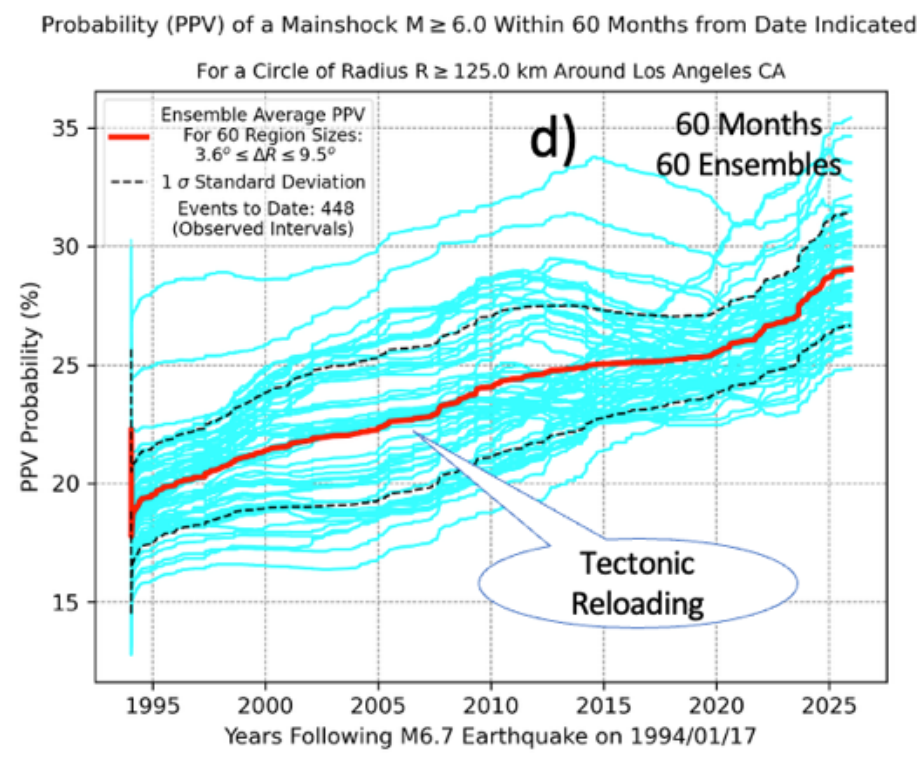

**Figure 5.** Validation of the ensemble method by comparing to the UCERF3 time-dependent forecast (Field et al., 2014) for the Los Angeles and San Francisco boxes defined in the UCERF3 forecast. The forecast curves at right begin at the time of the 1994 Northridge earthquake for the Los Angeles box, and the 1989 Loma Prieta earthquake for the San Francisco box, respectively. The current values of the Los Angeles and San Francisco ensemble 30-year forecasts are listed in the figure, together with the UCERF3 30-year forecasts.

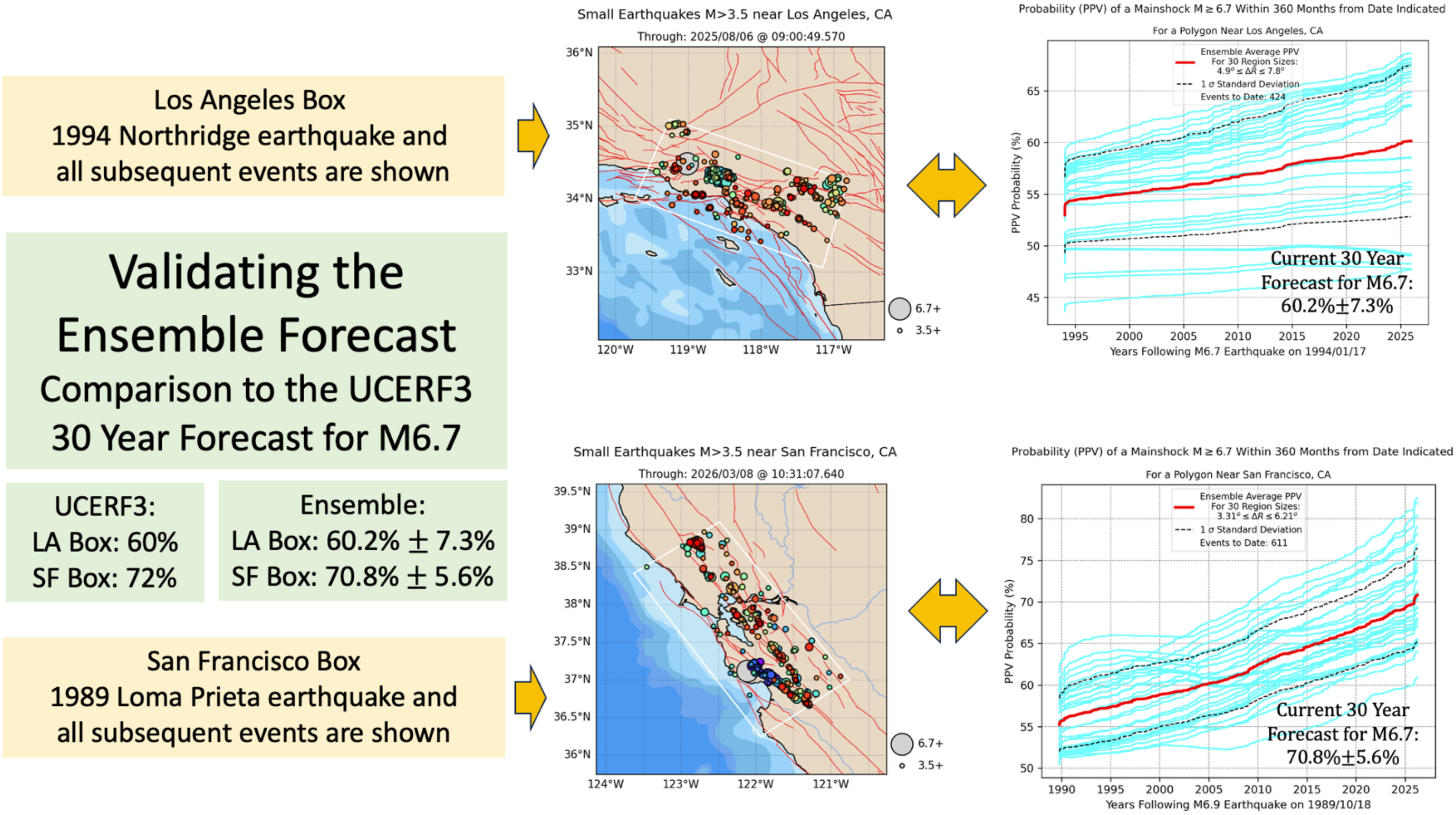